\documentstyle[preprint,tighten,eqsecnum,floats,epsfig,aps]{revtex}

\begin{document}

\draft

\title{ Ground state energy fluctuations in nuclear matter II}
\author{ A. De Pace$^a$, H. Feshbach$^b$ and A. Molinari$^a$ }
\address{
 ${}^{a}$ Dipartimento di Fisica Teorica dell'Universit\`a di Torino and \\
 Istituto Nazionale di Fisica Nucleare, Sezione di Torino, \\
 via P.Giuria 1, I-10125 Torino, Italy \\
 $^b$ Center for Theoretical Physics, \\
 Laboratory for Nuclear Science and Department of Physics, \\
 Massachusetts Institute of Technology, Cambridge, MA 02139, USA
}
\date{June 1999}

\maketitle

\begin{abstract}
Improvements are performed on a recently proposed statistical theory of the
mean field of a many-fermion system. The dependence of the predictions of the
theory upon its two basic ingredients, namely the Hartree-Fock energy and the
average energy of the two particle--two hole excitations, is explored.
\end{abstract}

\section{ Introduction }
\label{sec:intro}

In a recent paper \cite{Car98} a statistical theory of the bound states of 
atomic nuclei has been developed, where a new definition of the nuclear mean 
field (MF) and of the fluctuations around it (the ``error'') are given.

The aim of this theory is to provide an alternative route to overcome the 
problem presented by the strong forces acting among
the nucleons inside nuclei at short distances. These indeed entail very large
matrix elements between the ground and the highly excited nuclear states, thus
mixing into the former quite complex configurations. For this physics, in spite
of the many approaches worked out in the past, perturbative treatments are of
little value.

Hence, the MF is obtained in \cite{Car98} through an {\em energy average} that
smoothes out short time events. Furthermore, the matrix elements of the 
residual interaction (the one remaining after the {\em average} interaction is 
removed) are assumed to be {\em random} with vanishing average value.
These two elements are sufficient to obtain the MF and the ``error'', which of
course should be small for the MF to be meaningful.

Concerning the error, while its average is zero, the {\em average of its
square} is not and, in our framework, it can be computed. Indeed, we have
organized the energy fluctuations in classes of excited states of increasing 
complexity, each class yielding a contribution to the error. Notably, a 
criterion for the convergence of the expansion of the latter in the complexity
of the states has been obtained.

The above outlined statistical framework has been applied in \cite{Car98} to 
the problem of nuclear matter, in order to provide an appreciation of the 
results the approach leads to. In carrying out this analysis, however, 
rough approximations were made, which might cast some doubts on the
validity of our findings. Prominent among the latter is the one related to the
dramatic reduction of the strength of the residual interaction, with respect to
the bare one. Also the very fact that a statistical computation of the ground 
state energy of nuclear matter yields a reasonable result with a spectroscopic
factor varying between consistent values across the energy range of
the fluctuation is relevant.

In this paper we remove the approximations referred to above and
explore whether our previous findings survive the test of a more accurate
analysis. At the same time we examine the dependence upon the two inputs
required by our scheme, namely the Hartree-Fock (HF) energy and the average
energy of the two particle--two hole (2p-2h) excitations. Of significance is
the following: If a reliable estimate of the latter is available, then our 
theory essentially becomes parameter free. Furthermore, the closer the HF 
energy comes to the empirical binding energy of nuclear matter, the smaller the
fluctuations are. 

\section{ Basic formalism }
\label{sec:basic}

While for the general presentation of the formalism we refer the
reader to ref.~\cite{Car98}, here we confine ourselves to shortly revisit
the basic equations we shall need in the following. These stem from
the theory of nuclear reactions developed by one of us \cite{Fes92}, where the 
projection operators $P$ and $Q$ are introduced and the Schroedinger equation 
is split into two coupled equations
\begin{mathletters}
  \label{eq:Schr}
\begin{eqnarray}
  (E-H_{PP})(P\psi) &=& H_{PQ}(Q\psi) \\
  (E-H_{QQ})(Q\psi) &=& H_{QP}(P\psi) ,
\end{eqnarray}
\end{mathletters}
with obvious meaning of the symbols. From these the equation
\begin{equation}
  \label{eq:Hcal}
  {\cal H}(P\psi) = \left(H_{PP}+H_{PQ}
    \frac{1}{\left(\frac{\displaystyle 1}{\displaystyle e_Q}\right)^{-1}+
    W_{QQ}} H_{QP}\right)(P\psi) = E (P\psi)
\end{equation}
for the system's wave function projected into the $P$-space follows.
In (\ref{eq:Hcal}) the standard definitions 
\begin{mathletters}
\begin{eqnarray}
  e_Q    &=& E - H_{QQ} - W_{QQ} \\
  W_{QQ} &=& H_{QP}\frac{1}{E-H_{PP}}H_{PQ}
\end{eqnarray}
\end{mathletters}
have been adopted.

The strategy of our approach is first to implement the above formalism
by splitting the Hilbert space in a
$P$ sector embodying all the ``gentle'' physics and in a $Q$ sector
where the rapidly time dependent physics associated with the violent
two-body short range collisions among nucleons is inserted. 
Clearly, an exact prescription for such a partition cannot be given: This,
however, might be more an advantage than a flaw, since it adds to the
flexibility of the method.

Next, an energy averaging procedure is performed into the $Q$ -space: 
Specifically, the fast energy-dependent operator $1/e_Q$ is
averaged, leaving the weakly energy-dependent $W_{QQ}$ as it stands.

Thus, the equation 
\begin{equation}
  \label{eq:Hbar}
  \bar{\cal H}\langle P\psi\rangle = \left(H_{PP}+H_{PQ}
    \frac{1}{\left\langle\frac{\displaystyle 1}
    {\displaystyle e_Q}\right\rangle^{-1}+
    W_{QQ}} H_{QP}\right)\langle P\psi\rangle = \bar{E}_0 \langle P\psi\rangle
\end{equation}
is obtained. In (\ref{eq:Hbar}), $\bar{E}_0$ should be viewed as the
MF energy and the angle brackets mean energy and phase averaging.
Energy averaging is performed with the smoothing function
\begin{equation}
  \label{eq:rho}
  \rho(E,\bar{E}_0) = \frac{1}{2\pi i}\frac{1}{E-(\bar{E}_0-\epsilon)} ,
\end{equation}
$\epsilon$ being a real, positive parameter, which moves the poles of the 
$Q$-space away from the $P$-space (the more so, the larger $\epsilon$ is), thus
quenching the influence of the former over the latter.

With (\ref{eq:rho}) one obtains, by applying the Cauchy theorem to a contour
going along the real axis with a small semi-circle described positively around
the $\bar{E}_0-\epsilon$ pole,
\begin{eqnarray}
  \label{eq:eQav}
  \left\langle\frac{1}{e_Q}\right\rangle &=& \frac{1}{2\pi i}
    \int_{-\infty}^{\infty} dE\frac{1}{E-(\bar{E}_0-\epsilon)}
    \frac{1}{E-H_{QQ}-W_{QQ}(E)} \nonumber \\
  &=& \frac{1}{\bar{E}_0-\epsilon-H_{QQ}-W_{QQ}(E=\bar{E}_0-\epsilon)}
    \cong \frac{1}{\bar{E}_0-\epsilon-H_{QQ}-W_{QQ}(E=\bar{E}_0)} .
\end{eqnarray}
In performing the integration in the complex energy plane, account has been
taken of the position of the poles associated with the operator $1/e_Q$ (they
are located in the Im$E<0$ half-plane) \cite{Bro67}.

Furthermore, the approximation performed in (\ref{eq:eQav}) exploits the mild
energy dependence of $W_{QQ}$, but of course it requires, to be valid, a {\em
not too large} value of the parameter $\epsilon$. On the other hand, the latter
cannot be too small in order to lessen the impact of the $Q$-space on the $P$
one. Hence, $\epsilon$ is constrained to vary in a rather restricted range.

According to the above, we obtain for the $P$-space energy averaged \
Hamiltonian $\bar{\cal H}$ the expression
\begin{equation}
  \label{eq:Hbarav}
  \bar{\cal H} = H_{PP} + V_{PQ} V_{QP} \frac{1}{\bar{E}_0-\epsilon-E} ,
\end{equation}
the associated equation for the averaged wave function reading 
\begin{equation}
  \label{eq:Ppsiav}
  \bar{\cal H} \langle P\psi\rangle = \bar{E}_0\langle P\psi\rangle .
\end{equation}
The energy $\bar{E}_0$ in (\ref{eq:Ppsiav}) is commonly referred to as the MF
ground state energy. In (\ref{eq:Hbarav}), in analogy with ref.~\cite{Kaw73}, the
coupling operators
\begin{mathletters}
  \label{eq:VPQP}
\begin{equation}
  V_{PQ} = H_{PQ}\sqrt{\frac{\bar{E}_0-\epsilon-E}{\bar{E}_0-\epsilon-H_{QQ}}}
\end{equation}
and
\begin{equation}
  V_{QP} = \sqrt{\frac{\bar{E}_0-\epsilon-E}{\bar{E}_0-\epsilon-H_{QQ}}}H_{QP}
\end{equation}
\end{mathletters}
have been introduced. They should be viewed as representing what is left out of
the interaction among the nucleons after the average value of the latter has
been removed. We note that $V$ is zero when $\epsilon$ vanishes.

The above definitions are convenient because in terms of them the system
(\ref{eq:Schr}) can then be {\em exactly} recast as follows:
\begin{mathletters}
  \label{eq:Schrav}
\begin{eqnarray}
  (E-\bar{\cal H})(P\psi) &=& V_{PQ}(Q\psi) \\
  (E-H_{QQ})(Q\psi)       &=& V_{QP}(P\psi) .
\end{eqnarray}
\end{mathletters}
As shown in \cite{Car98}, is
\begin{equation}
  \label{eq:E-E0}
  E-\bar{E}_0 = \langle\phi_0|V_{PQ}\frac{1}{E-h_{QQ}}V_{QP}|\phi_0\rangle
    \equiv \Delta E ,
\end{equation}
with
\begin{eqnarray}
  \label{eq:hQQ}
  h_{QQ} &=& H_{QQ} + V_{QP}\left(\frac{1}{E-\bar{\cal H}}\right)^\prime V_{PQ}
    \nonumber \\
         &\equiv& H_{QQ} + \bar{W}_{QQ} .
\end{eqnarray}
In (\ref{eq:hQQ}) the prime implies the dropping of the $|\phi_0\rangle$
component of $1/(E-\bar{\cal H})$.

\section{ Energy fluctuation }
\label{sec:enefluc}

The physical significance of (\ref{eq:E-E0}) relates to the statistical
fluctuations of the energy around the average value $\bar{E}_0$ (the MF
energy). In fact, the average over the random phases of the wave-functions of 
the $Q$-space leads to
\begin{equation}
  \label{eq:DE0}
  \langle\Delta E\rangle = 0 .
\end{equation}
To implement the above constraint we split $\Delta E$ according to
\begin{equation}
  \label{eq:DEm}
  \Delta E = \sum_{m=1}^r \Delta E_m ,
\end{equation}
with
\begin{eqnarray}
  \Delta E_m &=& \langle\phi_0|V_{01}G_1 V_{12}\dots V_{m-1,m}G_m V_{m,m-1}
    \dots V_{10}|\phi_0\rangle , \\ \\
  G_k        &=& \frac{1}{E-h_{QQ}-V_{k,k+1}G_{k+1}V_{k+1,k}}
\end{eqnarray}
and
\begin{equation}
  V_{ij} = Q_i V Q_j .
\end{equation}
Actually, in (\ref{eq:DEm}) the index $m$ labels different sets of states of 
increasing complexity, which contribute {\em independently} to the error, the 
number of sets being $r$. Moreover, for each $m$
\begin{equation}
  \label{eq:DEm0}
  \langle\Delta E_m\rangle = 0 ;
\end{equation}
hence (\ref{eq:DE0}) follows.

We shall confine ourselves to consider the first set ($m=1$) only and will 
enforce (\ref{eq:DEm0}) by defining 
\begin{equation}
  \Delta E_1 = \langle\phi_0|V_{01}G_1 V_{10}|\phi_0\rangle -
    \left[\langle\phi_0|V_{01}G_1 V_{10}|\phi_0\rangle\right]_{\text AV} ,
\end{equation}
whose square we shall evaluate in the following.

Now, while $\langle\Delta E_m\rangle$ vanishes, $\langle(\Delta E_m)^2\rangle$
does not. We compute it, again assuming random matrix elements, as in
\cite{Car98}, but dropping a severe assumption there made. For this purpose let
the ground state, either of nuclear matter or of a finite nucleus, be $\phi_0$.
Assuming the residual interaction to be of two-body character, then the $m=1$ 
set will be made up of 2p-2h states $\psi_{2\beta}$ and the propagator of the
system in the $m=1$ sector will obey the equation
\begin{equation}
  G_1^{-1}\psi_{2\beta} = (E-h_{11})\psi_{2\beta} 
    = (E-\epsilon_{2\beta})\psi_{2\beta} ,
\end{equation}
the index $\beta$ embodying the quantum numbers identifying the 2p-2h states.
Then
\begin{eqnarray}
  \label{eq:dE1ex}
  (\delta E_1)^2 &\equiv& \biglb\langle(\Delta E_1)^2\bigrb\rangle - 
    \biglb\langle\Delta E_1\bigrb\rangle^2 \\
                 &=& \Bigglb\langle\sum_{\beta\gamma}
    \frac{|\langle\psi_{2\beta}|V|\phi_0\rangle|^2 
          |\langle\psi_{2\gamma}|V|\phi_0\rangle|^2}
         {(E-\epsilon_{2\beta})(E-\epsilon_{2\gamma})} \Biggrb\rangle -
    \Bigglb\langle\sum_\beta
    \frac{|\langle\psi_{2\beta}|V|\phi_0\rangle|^2}
         {(E-\epsilon_{2\beta})} \Biggrb\rangle^2 \nonumber \\
                 &\cong& \frac{1}{(E-\bar{\epsilon}_2)^2}
    \left\{\Bigglb\langle\sum_{\beta\gamma}
      \langle\psi_{2\beta}|V|\phi_0\rangle
      \langle\psi_{2\beta}|V|\phi_0\rangle^*
      \langle\psi_{2\gamma}|V|\phi_0\rangle
      \langle\psi_{2\gamma}|V|\phi_0\rangle^*
    \Biggrb\rangle \right. \nonumber \\
                 && \qquad\qquad\left.-\Bigglb\langle\sum_\beta
    |\langle\psi_{2\beta}|V|\phi_0\rangle|^2\Biggrb\rangle^2\right\}
    \nonumber \\
                 &=& \frac{1}{(E-\bar{\epsilon}_2)^2} \left\{
    \Bigglb\langle
      \sum_\beta|\langle\psi_{2\beta}|V|\phi_0\rangle|^2\Biggrb\rangle
      \Bigglb\langle
      \sum_\gamma|\langle\psi_{2\gamma}|V|\phi_0\rangle|^2\Biggrb\rangle
    \right. \nonumber \\
                 && \qquad\qquad + \sum_{\beta\gamma}
      \Biglb\langle
      \langle\psi_{2\beta}|V|\phi_0\rangle^*
      \langle\psi_{2\gamma}|V|\phi_0\rangle
      \Bigrb\rangle
      \Biglb\langle
      \langle\psi_{2\beta}|V|\phi_0\rangle
      \langle\psi_{2\gamma}|V|\phi_0\rangle^*
      \Bigrb\rangle \nonumber \\
                 && \qquad\qquad \left. + \text{ fourth order terms } 
      -\Bigglb\langle\sum_\beta
        |\langle\psi_{2\beta}|V|\phi_0\rangle|^2\Biggrb\rangle^2\right\} ,
      \nonumber
\end{eqnarray}
where an average 2p-2h eigenvalue $\bar{\epsilon}_2$ has been introduced in the
denominator, which has not been averaged owing to its smooth energy dependence.

Next, neglecting in the above the fourth order and because of the randomness of
the phases of the states, we finally arrive at the expression 
\begin{equation}
  \label{eq:dE1appr}
  (\delta E_1)^2 \cong \frac{1}{(E-\bar{\epsilon}_2)^2}\sum_\beta
    \Biglb\langle|\langle\psi_{2\beta}|V|\phi_0\rangle|^2\Bigrb\rangle^2 ,
\end{equation}
where the sum goes over the 2p-2h states contained in the energy range
$\overline{\Delta E}_1\cong\epsilon$ around $\bar{\epsilon}_2$ over which the 
average is carried out.

\section{ A simple projection operator }
\label{eq:projop}

In order to apply the formalism previously outlined a choice for the projection
operator $P$ should be made. We start with the simplest one, namely we take
\begin{equation}
  \label{eq:PHF}
  P = |\chi_{\text HF}\rangle\langle\chi_{\text HF}| ,
\end{equation}
$|\chi_{\text HF}\rangle$ being the HF ground state wave-function
(the Fermi sphere in the nuclear matter case). 

Then, by combining (\ref{eq:PHF}) with eq.~(\ref{eq:Ppsiav}), it is not 
difficult to derive the MF equation (see ref.~\cite{Car98})
\begin{equation}
  \label{eq:MFeq}
  \bar{E}_0 = E_{\text HF} + \frac{\beta^2}{\bar{E}_0-\epsilon-E} ,
\end{equation}
which relates the MF energy $\bar{E}_0$, the HF energy $E_{\text HF}$ and the
true energy of the system $E$. In (\ref{eq:MFeq})
\begin{equation}
  \label{eq:beta2}
  \beta^2 = \sum_{\text 2p-2h}
    |\langle\psi_{\text 2p-2h}|V|\chi_{\text HF}\rangle|^2 .
\end{equation}
To proceed further, we elaborate the relation (\ref{eq:dE1appr}) in order to
express the fluctuations of the ground state energy around $\bar{E}_0$ also in
terms of $\beta^2$. For this purpose, we write
\begin{eqnarray}
  \label{eq:dE1ar}
  (\delta E_1)^2 &=& \frac{1}{(E-\bar{\epsilon}_2)^2} \sum_{\beta\gamma}
    \Biglb\langle 
    \langle\psi_{2\beta}|V|\phi_0\rangle^*
    \langle\psi_{2\gamma}|V|\phi_0\rangle
    \Bigrb\rangle 
    \Biglb\langle 
    \langle\psi_{2\beta}|V|\phi_0\rangle
    \langle\psi_{2\gamma}|V|\phi_0\rangle^*
    \Bigrb\rangle \nonumber \\
  &=& \frac{1}{(E-\bar{\epsilon}_2)^2} 
    {\cal A}\left\{ \sum_{\beta\gamma}
    \Biglb\langle 
    \langle\psi_{2\beta}|V|\phi_0\rangle^*
    \langle\psi_{2\gamma}|V|\phi_0\rangle
    \Bigrb\rangle \right\} \sum_{\beta\gamma}
    \Biglb\langle 
    \langle\psi_{2\beta}|V|\phi_0\rangle
    \langle\psi_{2\gamma}|V|\phi_0\rangle^*
    \Bigrb\rangle \nonumber \\
  && + {\cal A}\left\{ \sum_{\beta\gamma}
    \Biglb\langle 
    \langle\psi_{2\beta}|V|\phi_0\rangle
    \langle\psi_{2\gamma}|V|\phi_0\rangle^*
    \Bigrb\rangle \right\} \sum_{\beta\gamma}
    \Biglb\langle 
    \langle\psi_{2\beta}|V|\phi_0\rangle^*
    \langle\psi_{2\gamma}|V|\phi_0\rangle
    \Bigrb\rangle \nonumber \\ 
  &=& \frac{2}{(E-\bar{\epsilon}_2)^2}
    {\cal A}\left\{ \sum_{\beta\gamma}
    \Biglb\langle 
    \langle\psi_{2\beta}|V|\phi_0\rangle^*
    \langle\psi_{2\gamma}|V|\phi_0\rangle
    \Bigrb\rangle \right\} \sum_{\beta\gamma}
    \Biglb\langle 
    \langle\psi_{2\beta}|V|\phi_0\rangle^*
    \langle\psi_{2\gamma}|V|\phi_0\rangle
    \Bigrb\rangle .
\end{eqnarray}
In the above, the operator ${\cal A}$ corresponds to the arithmetic mean of the
quantity within the angle brackets over the set of the 2p-2h states in the
energy interval $\epsilon$ centered around $\bar{\epsilon}_2$. 
Moreover, because of the randomness of the wave functions, the sum over 
$\beta\ne\gamma$ vanishes. We thus obtain
\begin{eqnarray}
  {\cal A}\left\{ \sum_{\beta\gamma}
    \Biglb\langle 
    \langle\psi_{2\beta}|V|\phi_0\rangle^*
    \langle\psi_{2\gamma}|V|\phi_0\rangle
    \Bigrb\rangle \right\} &\cong& \frac{1}{\cal N} 
    \sum_\gamma \Biglb\langle |\langle\psi_{2\gamma}|V|\phi_0\rangle|^2
    \Bigrb\rangle \nonumber \\
  &\cong& \frac{1}{\cal N} \beta^2 ,
\end{eqnarray}
${\cal N}$ being the number of states over which the sum is performed.

To obtain the latter we make use of Ericson's statistical theory of nuclear
states \cite{Eri60}, which yields for the density of the spin $J$ nuclear 
levels set up with $N$ particle (p)-hole (h) ($N$=p+h) excitations the 
expression
\begin{equation}
  \label{eq:rhoN}
  \rho^{(N)}_{ph}({\cal E},J) = \frac{g(g{\cal E})^{N-1}}{p!h!(N-1)!}
    \frac{2J+1}{\sqrt{8\pi}\sigma^3 N^{3/2}} \exp[-(2J+1)^2/(8N\sigma^2)] ,
\end{equation}
where
\begin{equation}
  \label{eq:sigma}
  \sigma^2 = {\cal F}\sqrt{\frac{\cal E}{a}}\frac{1}{\hbar^2}
\end{equation}
is the so-called spin cut-off parameter. In (\ref{eq:sigma}) $a$ is a constant
and ${\cal F}$ the system's moment of inertia. Note that the energy ${\cal E}$
in (\ref{eq:rhoN}) is the positive excitation energy of the states of concern.

The number of levels is then given by the expression 
\begin{equation}
  {\cal N} = \frac{\overline{\Delta E}_1}{{\cal D}_2} 
    \cong \epsilon \rho^{(4)}_{ph}({\cal E},0) ,
\end{equation}
${\cal D}_2$ being the distance in energy between the spin-zero 2p-2h
excitations, -- the ones we confine ourselves, for the sake of simplicity, to
consider. Hence, from (\ref{eq:dE1ar}), it follows 
\begin{equation}
  \biglb\langle(E-\bar{E}_0)^2\bigrb\rangle = \frac{2}{(E-\bar{\epsilon}_2)^2}
    \frac{\beta^4}{\cal N}
\end{equation}
and, by taking the square root,
\begin{equation}
  \label{eq:flucteq}
  E-\bar{E}_0 = \pm \frac{1}{E-\bar{\epsilon}_2}\sqrt{\frac{2}{\cal N}}\beta^2.
\end{equation}
As in ref.~\cite{Car98}, eqs.~(\ref{eq:MFeq}) and (\ref{eq:flucteq}) together 
make up two non-linear systems, eq.~(\ref{eq:flucteq}) being taken with the 
plus and the minus sign, respectively, in the first and in the second system.
The difference with the formalism of ref.~\cite{Car98} lies in the 
factor $1/\sqrt{\cal N}$.

In our approach the unknown quantities in the systems are either $E$ and
$\bar{E}_0$, if a model for the matrix elements of the residual interaction is
available, or $\bar{E}_0$ and $\beta^2$, if $E$ is experimentally measured and
one wishes to extract the residual effective interaction from the data.
In this paper, we shall follow the second option.
For definiteness we call the solutions of the first system $\bar{E}_0^l$ and
$E^l$ (or $\beta^2_l$) and, likewise, the solutions of the second system 
$\bar{E}_0^u$ and $E^u$ (or $\beta^2_u$), the labels $l$ and $u$ 
referring to the lower and upper boundaries of the energy band embracing the
fluctuations of the ground state energy of the system.

Now, both systems lead to the quadratic equation 
\begin{equation}
  \bar{E}_0^2 - \bar{E}_0(E_{\text{HF}}+E+\epsilon) + 
    E_{\text{HF}}(E+\epsilon) - \beta^2 = 0
\end{equation}
for the mean field energy. We take the solution with the minus sign in front of
the square root, namely
\begin{equation}
  \bar{E}_0 = \frac{1}{2} \left\{(E_{\text{HF}}+E+\epsilon) -
    \sqrt{(E_{\text{HF}}-E-\epsilon)^2+4\beta^2} \right\} ,
\end{equation}
since then, in the limit of vanishing residual interaction, which entails the
vanishing of $\epsilon$ (see eqs.~(\ref{eq:VPQP})), we obtain
\begin{equation}
  \bar{E}_0 = E ,
\end{equation}
which is consistent with eq.~(\ref{eq:flucteq}) in the same limit. Then, 
when $\beta^2\to 0$ the fluctuation of the ground state energy vanishes and the
latter coincides with the mean field value.

Concerning the residual effective interaction, one gets from the first system
the expression 
\begin{eqnarray}
  \label{eq:b2l}
  \beta^2_l &=& \sqrt{{\cal N}/2} \frac{E_l-\alpha\epsilon}{2} \left\{ \left[
    E_l(1+\sqrt{{\cal N}/2})-\epsilon(1+\sqrt{{\cal N}/2}\alpha)-E_{\text{HF}} 
    \right]
    \right. \nonumber \\ 
  && \qquad \left.
    + \sqrt{ \left[E_l(1+\sqrt{{\cal N}/2})-\epsilon(1+\sqrt{{\cal N}/2}\alpha)
    - E_{\text{HF}}\right]^2 + 4\epsilon(E_l-E_{\text{HF}})} \right\}
\end{eqnarray}
and from the second system the expression
\begin{eqnarray}
  \label{eq:b2u}
  \beta^2_u &=& \sqrt{{\cal N}/2} \frac{E_u-\alpha\epsilon}{2} \left\{ \left[
  E_u(-1+\sqrt{{\cal N}/2})-\epsilon(-1+\sqrt{{\cal N}/2}\alpha)+E_{\text{HF}} 
    \right] \right. \nonumber \\
  && \qquad \left.
    +\sqrt{\left[E_u(-1+\sqrt{{\cal N}/2})-\epsilon(-1+\sqrt{{\cal N}/2}\alpha)
    + E_{\text{HF}}\right]^2 + 4\epsilon(E_u-E_{\text{HF}})} \right\} .
\end{eqnarray}
In the above, in conformity to the notations of ref.~\cite{Car98}, we have set 
\begin{equation}
  \label{eq:eps2}
  \bar{\epsilon}_2 \equiv \alpha\epsilon
\end{equation}
and the solutions with the minus sign in front of the square root have been
discarded because they correspond to a residual effective interaction
non-vanishing with $\epsilon$.
Clearly, if an estimate of the average energy of the 2p-2h states can be
derived, then $\alpha$ in (\ref{eq:eps2}) should not be viewed as a parameter,
but merely as a convenient tool to express $\bar{\epsilon}_2$. 

Our solutions should fulfill two constraints: The first one is
\begin{equation}
  \label{eq:E0ul}
  \bar{E}_0^u = \bar{E}_0^l ,
\end{equation}
namely the two mean fields stemming from the two systems should be identical.
The second relates to the spectroscopic factor $S$ defined through the
normalization of the ground state wave-function projected onto the $P$-space,
namely
\begin{equation}
  \langle P\psi|P\psi\rangle = S^2 ,
\end{equation}
and found to be
\begin{equation}
  \label{eq:S2}
  S^2 = \left\{ \frac{3}{2} + \frac{1}{2} 
    \frac{E_{\text{HF}} - E - \epsilon - 2d\beta^2/dE}
    {\sqrt{(E_{\text{HF}} - E - \epsilon)^2 + 4\beta^2}} + 
    \frac{\epsilon}{\bar{E}_0 - E - \epsilon} \right\}^{-1}
\end{equation}
in ref.~\cite{Car98}.
We require the values of $S^2$, which approaches one (as it should) when 
$\beta^2\to 0$, to be ``reasonable'' when computed with our solutions for $E$ 
and $\beta^2$ on the upper and lower bounds of the energy band, respectively.
Actually, as it is apparent from (\ref{eq:S2}), $S^2$ depends also upon the
energy derivative of the residual effective interaction.
This reads
\begin{eqnarray}
  \frac{d\beta^2_l}{dE} &=& \frac{\sqrt{{\cal N}/2}}{2}\Biggl\{
    E_l(1+\sqrt{{\cal N}/2})-\epsilon(1+\sqrt{{\cal N}/2}\alpha)-E_{\text{HF}}
    \nonumber \\
  && \quad + 
    \sqrt{ \left[E_l(1+\sqrt{{\cal N}/2})-\epsilon(1+\sqrt{{\cal N}/2}\alpha)
    - E_{\text{HF}}\right]^2 + 4\epsilon(E_l-E_{\text{HF}})} \nonumber \\
  && \quad + (E_l - \alpha\epsilon) (1+\sqrt{{\cal N}/2}) \nonumber \\
  && \quad \times \Biggl[
    1+\frac{E_l(1+\sqrt{{\cal N}/2})-\epsilon(1+\sqrt{{\cal N}/2}\alpha)
    -E_{\text{HF}}+2\epsilon/(1+\sqrt{{\cal N}/2})}{\sqrt{[
    E_l(1+\sqrt{{\cal N}/2})-\epsilon(1+\sqrt{{\cal N}/2}\alpha)
    -E_{\text{HF}}]^2 + 4\epsilon(E_l-E_{\text{HF}})}} \Biggr]\Biggr\}
\end{eqnarray}
on the lower bound and
\begin{eqnarray}
  \frac{d\beta^2_u}{dE} &=& \frac{\sqrt{{\cal N}/2}}{2}\Biggl\{
    E_u(-1+\sqrt{{\cal N}/2})-\epsilon(-1+\sqrt{{\cal N}/2}\alpha)
    +E_{\text{HF}} \nonumber \\
  && \quad + 
    \sqrt{\left[E_u(-1+\sqrt{{\cal N}/2})-\epsilon(-1+\sqrt{{\cal N}/2}\alpha)
    + E_{\text{HF}}\right]^2 + 4\epsilon(E_u-E_{\text{HF}})} \nonumber \\
  && \quad + (E_u-\alpha\epsilon) \nonumber \\
  && \quad \times \Biggl[\sqrt{{\cal N}/2}-1+\frac{
    (E_u(-1+\sqrt{{\cal N}/2})-\epsilon(-1+\sqrt{{\cal N}/2}\alpha)
    +E_{\text{HF}})(\sqrt{{\cal N}/2}-1) + 2 \epsilon}{\sqrt{
    [E_u(-1+\sqrt{{\cal N}/2})-\epsilon(-1+\sqrt{{\cal N}/2}\alpha)
    +E_{\text{HF}}]^2 + 4\epsilon(E_u-E_{\text{HF}})}} \Biggr]\Biggr\}
   \nonumber \\
\end{eqnarray}
on the upper one and reduces to the corresponding expression found in 
ref.~\cite{Car98} when ${\cal N}\to2$.

\section{ Nuclear matter }
\label{eq:nuclmatt}

In this section we compute the nuclear matter mean field energy $\bar{E}_0$
together with its error. In addition we obtain, by solving the systems of 
eqs.~(\ref{eq:MFeq}) and (\ref{eq:flucteq}), the lower ($\beta^2_l$) and the
upper ($\beta^2_u$) residual effective interaction.

To carry out this task we exploit the empirical nuclear matter energy 
\begin{equation}
  \label{eq:ENM}
  E_{\text{NM}} = [-16+39.5(k_F-1.36)^2]\,\text{MeV} ,
\end{equation}
with $k_F$ in fm$^{-1}$.
Notice that formula (\ref{eq:ENM}) embodies not only the experimental binding
energy, but the saturation density and the compression modulus of nuclear
matter as well. Furthermore, to start with we take $E_{\text{HF}}$ as given in
ref.~\cite{Car98}, where a simple, four parameter two-body force with 
short-range repulsion followed by intermediate-range attraction was found to 
yield a minimum for the HF energy of -7.3 MeV at $k_F=1.78$~fm$^{-1}$ for some 
choice of the parameters.

Concerning ${\cal N}$ we assume, for a nucleus of mass number $A$, the
following values for the quantities entering into its definition:
\begin{mathletters}
\begin{eqnarray}
  a &=& \frac{A}{8}\ \text{MeV}^{-1} \\
  \label{eq:Fcal}
  {\cal F} &=& \frac{2}{3}m_N\int d\bbox{r}\, r^2\rho(r) = 
    \frac{8\pi}{15}m_N\rho_0 R^5 \\
  g &=& \frac{3}{2}\frac{A}{\epsilon_F} .
\end{eqnarray}
\end{mathletters}
In the above, (\ref{eq:Fcal}) corresponds to the moment of inertia of a
constant mass distribution $\rho(r) = \rho_0 = 2k_F^3/3\pi^2$ of radius 
$R=r_0 A^{1/3}$ ($r_0=1.2$ fm), $m_N$ being the nucleon's mass, and the
Fermi energy $\epsilon_F$ is computed with the standard nuclear matter value of
the Fermi wave number, i.~e. $k_F=1.36$ fm$^{-1}$.

Inserting the above in (\ref{eq:rhoN}) we get, for a $A=208$ nucleus taken as a
good representative of nuclear matter, 
\begin{equation}
  \rho_{2p-2h}^{(4)}({\cal E},J=0) = 3.76\cdot 10^{-2}{\cal E}^{9/4}
    \exp(-0.0013/\sqrt{{\cal E}}) \ \text{MeV}^{-1} ,
\end{equation}
${\cal E}=\alpha\epsilon-E_{\text{NM}}$ being the excitation energy of the
2p-2h states. 

Next, by introducing the fluctuation energy $W$ and setting accordingly
\begin{mathletters}
\begin{equation}
  E_l = E_{\text{NM}}-\frac{W}{2}
\end{equation}
and
\begin{equation}
  E_u = E_{\text{NM}}+\frac{W}{2}
\end{equation}
\end{mathletters}
we obtain all the ingredients required to get, from (\ref{eq:b2l}) and
(\ref{eq:b2u}), $\beta^2_l$ and $\beta^2_u$ as a function of $\epsilon$ and
$\alpha$ for given values of $W$ and ${\cal E}$.

With these the equal mean field condition (\ref{eq:E0ul}) can be cast into the
form of the following equation
\begin{equation}
  \label{eq:ElEu}
  E_l-\sqrt{(E_{\text{HF}}-E_l-\epsilon)^2+4\beta^2_l} =
    E_u-\sqrt{(E_{\text{HF}}-E_u-\epsilon)^2+4\beta^2_u} ,
\end{equation}
which we have numerically solved for $W=1$, 2 and 3 MeV and for ${\cal E}=14$
MeV. The latter value entails $\bar{\epsilon}_2\equiv\alpha\epsilon=-2$ MeV and
corresponds to ${\cal E}\cong2\hbar\omega$, $\hbar\omega=41/A^{1/3}$ ($\cong7$
MeV for $A=208$) being the parameter of the harmonic oscillator well modelling
the shell model binding potential.

\begin{figure}[p]
\begin{center}
\mbox{\epsfig{file=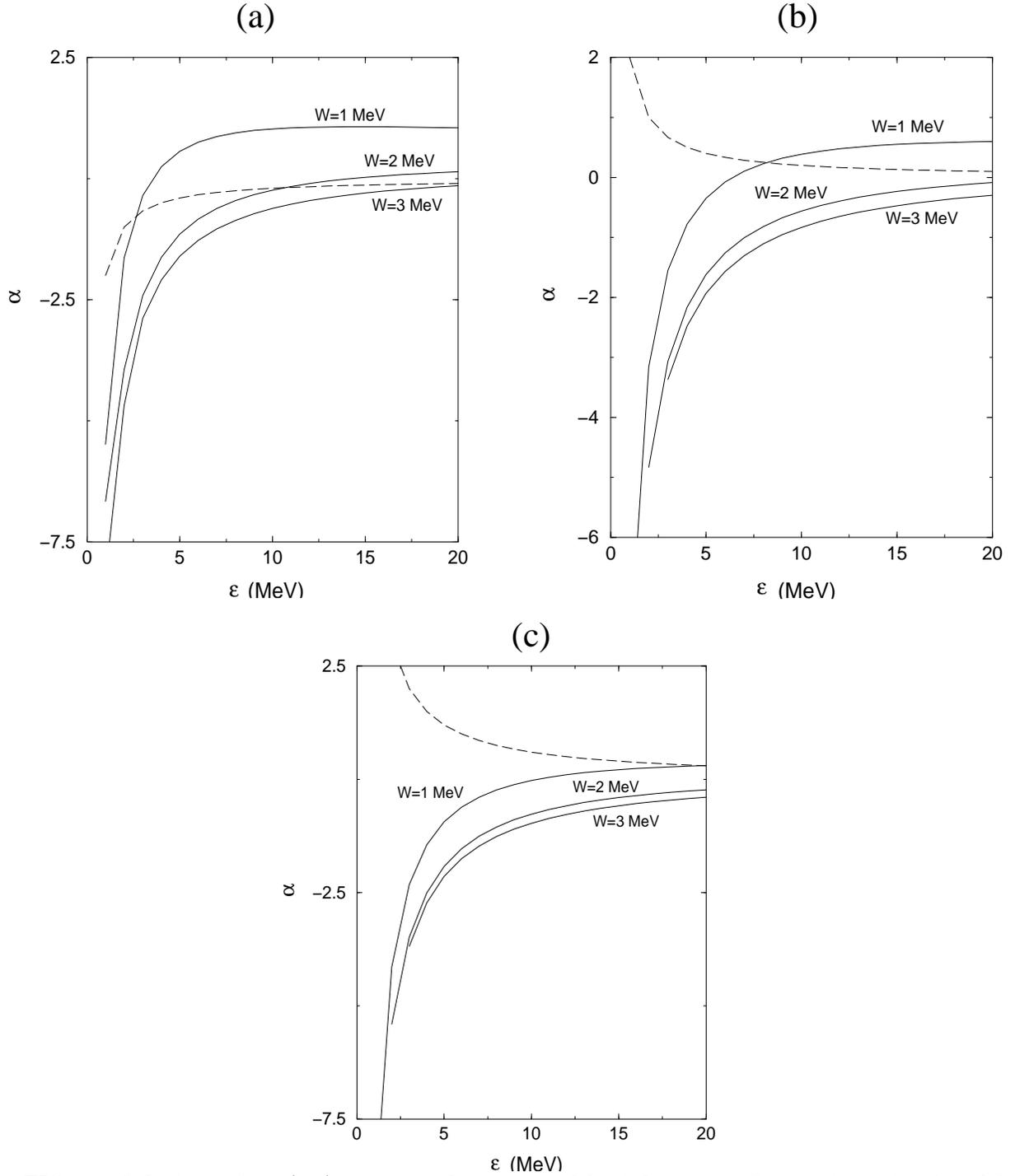}}
\caption{ Solutions of eq.~(\protect\ref{eq:ElEu}) corresponding to a width of
  the energy fluctuations of 1, 2 and 3 MeV (solid lines); the dashed lines
  correspond to $\alpha\epsilon\equiv\bar{\epsilon}_2=E_{\text{NM}}+{\cal E}$:
  (a) ${\cal E}=14$ MeV; (b) ${\cal E}=18$ MeV; (c) ${\cal E}=22$ MeV. }
\label{fig:fig_alfeps}
\end{center}
\end{figure}
Our results are displayed in Fig.~\ref{fig:fig_alfeps}a, where a continuum set 
of solutions is seen to exist for each value of $W$.
Should the above estimate of ${\cal E}$ be adequate, then the intercepts of the
hyperbola $\alpha=-2/\epsilon$ with the equal mean field curves of
Fig.~\ref{fig:fig_alfeps} would set the values (one for each choice of $W$) of 
the sole parameter entering in our theory, namely $\epsilon$, thus providing 
the {\em consistent} solutions of our approach. 

Of course, the estimate ${\cal E}\cong14$ MeV is only orienting: Therefore,
we display as well in Figs.~\ref{fig:fig_alfeps}b and \ref{fig:fig_alfeps}c the
curves corresponding to the choices ${\cal E}=18$ ($\bar{\epsilon}_2=2$) MeV 
and ${\cal E}=22$ ($\bar{\epsilon}_2=6$) MeV. Note the dependence upon 
${\cal E}$ of the solutions of (\ref{eq:ElEu}).
Indeed, the larger ${\cal E}$ is, the smaller the energy fluctuation becomes,
as it should be expected from the fluctuation equation (\ref{eq:flucteq}).

\begin{table}[p]
\begin{center}
\begin{tabular}{cddddddd}
  $W$ & $\epsilon$ & $\alpha$ & 
  $\bar{E}_0^l\equiv\bar{E}_0^u$ & $S^l$ & $S^u$ & 
  $\beta^2_l$ & $\beta^2_u$ \\ 
  (MeV) & (MeV) & & (MeV) & & & (MeV$^2$/nucleon) & (MeV$^2$/nucleon) \\ 
\tableline
  1 & 2.56  & -0.78  & -16.11 & 1.11 & 0.91 & 24.12  & 35.25   \\ 
  2 & 10.62 & -0.19  & -16.16 & 1.06 & 0.95 & 109.35 & 131.70  \\ 
  3 & 23.42 & -0.085 & -16.25 & 1.04 & 0.96 & 249.75 & 283.56  \\
\end{tabular}
\caption{ Solutions of eq.~(\protect\ref{eq:ElEu}) and 
  $\alpha\epsilon=E_{\text{NM}}+{\cal E}$ for ${\cal E}=14$ MeV and for three
  values of the width $W$ of the energy fluctuations; there are also reported
  the corresponding values of mean field, spectroscopic factors and effective
  interactions. The bare potential used in the calculations yields a HF field 
  with a minimum of -7.3 MeV at $k_F=1.78$ fm$^{-1}$, a HF energy of -4.99
  MeV at $k_F=1.36$ fm$^{-1}$ and $\beta^2_{\text{bare}}=5.2\cdot10^4$
  MeV$^2$/nucleon. }
\label{tab:E14}
\vskip 15mm
\begin{tabular}{cddddddd}
  $W$ & $\epsilon$ & $\alpha$ & 
  $\bar{E}_0^l\equiv\bar{E}_0^u$ & $S^l$ & $S^u$ & 
  $\beta^2_l$ & $\beta^2_u$ \\ 
  (MeV) & (MeV) & & (MeV) & & & (MeV$^2$/nucleon) & (MeV$^2$/nucleon) \\ 
\tableline
  1 & 8.15  & 0.25  & -16.04 & 1.04 & 0.96 & 85.11  & 96.16   \\ 
  2 & 32.40 & 0.062 & -16.09 & 1.03 & 0.97 & 349.39 & 371.58  \\ 
  3 & 71.49 & 0.028 & -16.16 & 1.03 & 0.96 & 783.36 & 816.86  \\
\end{tabular}
\caption{ As in Table \protect\ref{tab:E14}, but for ${\cal E}=18$ MeV. }
\label{tab:E18}
\vskip 15mm
\begin{tabular}{cddddddd}
  $W$ & $\epsilon$ & $\alpha$ & 
  $\bar{E}_0^l\equiv\bar{E}_0^u$ & $S^l$ & $S^u$ & 
  $\beta^2_l$ & $\beta^2_u$ \\ 
  (MeV) & (MeV) & & (MeV) & & & (MeV$^2$/nucleon) & (MeV$^2$/nucleon) \\ 
\tableline
  1 & 19.78  & 0.30  & -16.02 & 1.02 & 0.98 & 213.03  & 224.06   \\ 
  2 & 78.46  & 0.076 & -16.05 & 1.03 & 0.97 & 858.10  & 880.24   \\ 
  3 & 173.85 & 0.035 & -16.12 & 1.03 & 0.96 & 1919.00 & 1952.38  \\
\end{tabular}
\caption{ As in Table \protect\ref{tab:E14}, but for ${\cal E}=22$ MeV. }
\label{tab:E22}
\vskip 15mm
\begin{tabular}{dddddddd}
  $W$ & $\epsilon$ & $\alpha$ & 
  $\bar{E}_0^l\equiv\bar{E}_0^u$ & $S^l$ & $S^u$ & 
  $\beta^2_l$ & $\beta^2_u$ \\ 
  (MeV) & (MeV) & & (MeV) & & & (MeV$^2$/nucleon) & (MeV$^2$/nucleon) \\ 
\tableline
  0.4 &  3.99 & -0.50  & -16.01 & 1.04 & 0.95 & 14.19 & 15.68  \\ 
  0.5 &  6.24 & -0.32  & -16.01 & 1.04 & 0.95 & 22.41 & 24.28  \\ 
    1 & 24.80 & -0.081 & -16.03 & 1.05 & 0.94 & 91.10 & 94.85  \\
\end{tabular}
\caption{ As in Table \protect\ref{tab:E14}, but with a bare potential yielding
  a HF field with a minimum of -13 MeV at $k_F=1.5$ fm$^{-1}$, a HF energy of
  -12.28 MeV at $k_F=1.36$ fm$^{-1}$. }
\label{tab:E14EHF}
\end{center}
\end{table}
Finally, we collect our results, namely the values of $\epsilon$ and $\alpha$
corresponding to the intercepts for $W=1$, 2 and 3 MeV, together with the
associated values of $\bar{E}_0^l$, $\bar{E}_0^u$, $S^l$, $S^u$, $\beta^2_l$
and $\beta^2_u$ for ${\cal E}=14$, 18 and 22 MeV in Tables \ref{tab:E14},
\ref{tab:E18} and \ref{tab:E22}, respectively.

These predictions of our theory should be examined with the guiding principle
of having a parameter $\epsilon$ neither too large, nor too small and a
consistent spectroscopic factor $S$.

From Tables \ref{tab:E14}, \ref{tab:E18} and \ref{tab:E22} we infer that the
present theory comes indeed close to fulfilling the above requirements for a
fluctuation energy, say, $W\lesssim1\div2$ MeV and an average excitation energy
of the 2p-2h states ${\cal E}<20$ MeV.

Notably, this outcome is reached with a major reduction of the residual
effective interaction with respect to the bare $\beta^2$, which, with the 
parameters characterizing the force taken as in \cite{Car98}, reads 
$\beta^2=5.2\cdot10^4$ MeV$^2$/nucleon. 
Thus the findings of ref.~\cite{Car98} concerning both the size of the error 
of the nuclear matter ground state energy and the quenching of the bare 
interaction appear now established on a firmer ground.

Concerning the spectroscopic factor, its value comes close to one, a result
consistent with our effective interaction which is too weak to affect the
occupation probability of the single-particle levels inside the Fermi sphere
(in \cite{Car98} a sign error in $d\beta^2/dE$ yielded quite small values for 
$S$). The discrepancy with current estimates, which yield $S\approx0.7$ 
\cite{6}, is likely to stem from the restricted $P$-space we have adopted. 
Work is currently in progress to shed light on this point (by enlarging the 
$P$-space) \cite{prog}.

Another important point, is whether our conclusions remain true for a different
choice of $E_{\text{HF}}$. Actually, the HF energy employed by us should rather
be viewed as a tool needed in our framework than as a quantity precisely
computed. Indeed, in order to calculate $E_{\text{HF}}$, one should start from
a realistic two-body force, deduce the associated $G$-matrix, thus accounting
for the deterministic aspect of the short-range nucleon-nucleon correlations
and then perform a HF (in fact, a Brueckner-Hartree-Fock) calculation.
This we have not done. We have however varied the parameters of our simple
two-body force to get $E_{\text{HF}}$ closer to the empirical energy, namely
getting a HF minimum of -13 MeV at $k_F=1.5$ fm$^{-1}$ with roughly the same
empirical compressibility of nuclear matter, thus yielding
$E_{\text{HF}}=-12.28$ MeV at $k_F=1.36$ fm$^{-1}$.

The results thus obtained are quoted in Table \ref{tab:E14EHF}. 
It is clearly apparent that the closer $E_{\text{HF}}$ is to the true value of
the energy, the thinner the fluctuations are.

\section{ Conclusions }
\label{sec:concl}

In the present research we improve upon the statistical treatment of the mean
field of the atomic nuclei and nuclear matter we have recently proposed
\cite{Car98}. Specifically, we deal more realistically with the equation
expressing the fluctuation of the energy of the ground state by exploiting the
Ericson's formula for the density of the 2p-2h states.
The latter in turn allows us to introduce a self-consistency condition for the
mean field energy, which correlates the value of the energy averaging parameter
$\epsilon$ with the average excitation energy ${\cal E}$ of the 2p-2h states.
Should the latter be known, then no parameter would be left in our theory.

Lacking such a knowledge, we have explored the dependence of our predictions
for the ground state energy fluctuation $W$ upon ${\cal E}$.
As expected, we have found that the larger ${\cal E}$ is, the more remote the
$Q$-space becomes: hence, the smaller the $W$ one gets. However, while 
${\cal E}$ on the one side cannot be smaller than $2\hbar\omega$, on the other
cannot be too large because its actual value relates to the average energy of
those 2p-2h states connected to the ground state of $\bar{\cal H}$ by the
residual force $V$, which is found to be very weak.

In principle also a large $\epsilon$ would push away the $Q$-space, but then
our theory, as previously shown, would not work any longer.
In fact, as seen in the figures, the solutions associated with a large
$\epsilon$ also correspond to a large $W$, which does not make any sense.
Rather, the trustable solution of our theory is the one corresponding to a $W$
of one, at most two, MeV and to an energy averaging parameter
$\epsilon\cong2\div3$ MeV, these outcomes being accompanied by a quenching of
the bare interaction of more than an order of magnitude.

The latter finding should not surprise: in fact, our residual interaction has
been dove-tailed to vanish when $\epsilon\to0$ and $\epsilon$ actually is about
$2\div3$ MeV. Since the physics associated with the $Q$-space extends up to
several hundred MeV, the strong quenching of $V_{PQ}$ ($V_{QP}$) with respect
to $H_{PQ}$ ($H_{QP}$) immediately follows.

The above results, however, are obtained with a HF energy yielding a poor
approximation for the true energy. A $E_{\text{HF}}$ closer to the latter
entails, as we have seen, a smaller $W$ (but not a smaller $\epsilon$). 
We argue accordingly that our theory provides an appreciation of the error of
any theory attempting to calculate the ground state energy of nuclei and, of
course, the nearer the theory comes to the true eigenvalue, the smaller the
associated error is.

Finally, the unrealistic, but consistent, value of our spectroscopic factor has
already been commented upon. Here we wish to remark that the better estimate of
$S$ we expect by enlarging the $P$-space and by allowing configurations more 
complex than the 2p-2h ones to contribute, should go in parallel to a better 
appreciation of the error as well.

\acknowledgements
This work has been partially supported by the INFN-MIT ``Bruno Rossi'' Exchange
Program.


\begin{references}

\bibitem{Car98}       R. Caracciolo, A. De Pace, H. Feshbach, and A. Molinari,
                      {\em Ann. Phys. (N.Y.)} {\bf 262} (1998), 105;
                      Proceedings of the ``International School of Physics
                      E. Fermi'', Course CXXXVIII ``Unfolding the Matter of
                      Nuclei'', edited by A. Molinari and R. A. Ricci, IOS
                      Press, Amsterdam (1999) p.1.; 
                      see also, H. Feshbach,
                      {\em Phys. Rep.\/} {\bf 264} (1996), 137.

\bibitem{Fes92}       H. Feshbach,
                      ``Theoretical Nuclear Physics: Nuclear Reactions'',
                      J.~Wiley \& Sons, New York (1992).

\bibitem{Bro67}       G. E. Brown,
                      ``Unified Theory of Nuclear Models and Forces'',
                      North-Holland, Amsterdam (1967) p.30.

\bibitem{Kaw73}       M.~Kawai, A.K.~Kerman and K.W. McVoy,
                      {\em Ann. Phys. \/}  (N.Y.) {\bf 75} (1973), 156.

\bibitem{Eri60}       T. E. O. Ericson,
                      {\em Phil. Mag. Suppl. \/} {\bf 9} (1960), 425.

\bibitem{6}           V. R. Pandharipande, I. Sick, and P. K. A. 
                      deWitt-Huberts, 
                      {\em Rev. Mod. Phys. \/} {\bf 69} (1997), 981.

\bibitem{prog}        A. De Pace, H. Feshbach, and A. Molinari,
                      work in progress.

\end{references}
\end{document}